\newcommand{\Msun}{\mbox{$M_{\odot}$}}
\newcommand\aj{{AJ\,}}%
\newcommand\araa{{ARA\&A\,}}%
\newcommand\apj{{ApJ\,}}%
\newcommand\apjs{{ApJS\,}}%
\newcommand\apss{{Ap\&SS\,}}%
\newcommand\aap{{A\&A\,}}%
\newcommand\aaps{{A\&AS\,}}%
\newcommand\mnras{{MNRAS\,}}%
\newcommand\pasj{{PASJ\,}}%
\newcommand\ssr{{Space. Sci. Rev.\,}}%
\newcommand\solphys{{Sol. Phys.\,}}%
\newcommand\memsai{{MSAIT\,}}%

\documentclass[graybox, natbib, footinfo]{svmult}

\usepackage{mathptmx}       
\usepackage{helvet}         
\usepackage{courier}        
\usepackage{type1cm}        
\usepackage{makeidx}         
\usepackage{graphicx}        
\usepackage{multicol}        
\usepackage[bottom]{footmisc}

\makeindex             

\begin{document}

\title*{Uncertainties in stellar evolution models: convective overshoot}
\author{Alessandro Bressan, L\'eo Girardi, Paola Marigo, Philip Rosenfield \& Jing Tang}
\authorrunning{Bressan et al.}
\institute{Alessandro Bressan, Jing Tang \at SISSA, via Bonomea 265, I-34136 Trieste, Italy, \email{sbressan@sissa.it, tang@sissa.it}
\and Paola Marigo, Philip Rosenfield \at Dipartimento di Fisica e Astronomia Galileo Galilei,
           Universit\`a di Padova, Vicolo dell'Osservatorio 3, I-35122 Padova, Italy, \email{paola.marigo@unipd.it, philip.rosenfield@unipd.it}
\and Leo Girardi, Osservatorio Astronomico di Padova, Vicolo dell'Osservatorio 5,
           I-35122 Padova, Italy, \email{leo.girardi@oapd.inaf.it}}
%
%
%
%
\maketitle

\abstract{In spite of the great effort made in the last decades
to improve our understanding of stellar evolution,
significant uncertainties remain due to our poor knowledge of some
complex physical processes that require an empirical calibration, such as
the efficiency of the interior mixing related to convective overshoot.
Here we review the impact of convective overshoot on the evolution
of stars during the main Hydrogen and Helium burning phases.}

\section{Introduction}
\label{sec:1}
Thanks to the efforts of many different groups in the last decades,
stellar evolution has now reached a high degree of
accuracy and completeness.
Indeed, it can now account for a variety of internal physical processes,
follow the most advanced phases and deal with different
chemical compositions, so that one could in principle reproduce
any stellar environment disclosed by the continuously advancing observational
facilities. At the same time observations themselves have become more
and more detailed,  even providing direct access to star interiors, like in the case of
asteroseismology, thus  posing a real challenge to theory.
In spite of these efforts,
some physical processes,  because of their complexity,
still suffer of large uncertainties.
These processes are crucial when dealing with  advanced evolutionary phases,
e.g. the Red Giant Branch (RGB) and the Asymptotic Giant Branch (AGB),
and  even more for stellar populations that are not well represented
in the solar vicinity or in our Galaxy.
From the theoretical point of view there are several long lasting questions that
still lack a definitive answer, such as the issue of convective energy transport and mixing
and that of the efficiency of the mass-loss phenomenon.
This review deals mainly with one of such questions, the effect of convective mixing
during the main phases of stellar evolution.
We will summarise the current theoretical situation with emphasis
on those phases  where the uncertainties become more critical.

We may expect that adding new dimensions to the HR diagram,
such as those provided  by asteroseismology, could allow the biggest improvements just
where the uncertainties are the largest.

\section{The new release of stellar evolutionary tracks with PARSEC}
\label{sec:2}
We begin with a brief  review of the new code developed in Padova, PARSEC
({\sl\,P}adova {T\sl\,R}ieste {\sl\,S}tellar {\sl\,E}volution {\sl\,C}ode),
with which we obtained many of the results presented here.
A detailed description can be found in \cite{Bressan_etal12}
and a few more recent updates in \cite{Bressan_etal13}.
The equation of state (EOS) is computed with the FreeEOS code
developed and updated over the years by A.W.~Irwin \footnote{http://freeeos.sourceforge.net/}.
Opacities in the high-temperature regime, $4.2 \le \log(T/{\rm K}) \le 8.7$, are
obtained from the Opacity Project At Livermore
(OPAL) team \citep{IglesiasRogers_96}
while, in the low-temperature regime,
$3.2 \le \log(T/{\rm K}) \le 4.1$, we use opacities generated with our
\AE SOPUS\footnote{http://stev.oapd.inaf.it/aesopus} code \citep{MarigoAringer_09}.
Conductive opacities are included following \cite{Itoh_etal08}.
The nuclear reaction network consists of the p-p chains, the CNO tri-cycle, the
Ne--Na and Mg--Al chains and the most important $\alpha$-capture
reactions, including the $\alpha$-n reactions.
The reaction rates and the corresponding $Q$-values
are taken from the recommended rates in the JINA reaclib database
\citep{Cyburt_etal10}.
Microscopic diffusion, applied to all the elements considered in the code in the
approximations that they are all fully ionized,
is included following \cite{Salasnich99} with diffusion coefficients
calculated following \cite{Thoul_etal94}.
The energy transport in the convective regions is described according
to the mixing-length theory of \cite{mlt}.
The mixing length parameter
is fixed by means of the solar model calibration and turns out to be $\alpha_{\rm MLT}=1.74$.
A remarkable difference with respect to other solar calibrations concern
the partition of metals for which we assume
the abundances taken from \cite{GrevesseSauval_98} but for the
species recently revised by \cite{Caffau_etal11}.
According to this abundance compilation, the present-day Sun's
metallicity is $Z_{\odot}= 0.01524$, intermediate between the
most recent estimates, e.g.  $Z_{\odot}= 0.0141$ of \cite{Lodders_etal09}
or $Z_{\odot}= 0.0134$ of \cite{Asplund_etal09},
and the previous value of $Z_{\odot}= 0.017$ by  \cite{GrevesseSauval_98}.
We remind here that the assumption of a different metallicity for the Sun
affects directly the calibration of the MLT parameter.

\section{Convective Overshoot on the Main Sequence}
Besides the choice of the microscopic physics, the evolution during and after
the main sequence is affected by mixing processes related to
the efficiency of convective core overshoot, atomic diffusion and rotation.
In PARSEC we adopt a maximum overshooting efficiency  $\Lambda_{\rm~max}=0.5$,
i.e.\ a moderate amount of overshooting (\citealt{Bertelli_etal94}, \citealt{Girardi_etal00}),
corresponding to about $0.25\,H_P$ of overshoot region {\em above} the unstable region as
commonly adopted by other authors.
In the transition region between stars with radiative cores and those with convective cores,
say between  $M_{\rm\,O1}\leq~M\leq~M_{\rm\,O2}$,
we assume that the overshooting efficiency $\Lambda_{\rm~c}$
increases linearly with mass from zero to the maximum value.
We define $M_{\rm\,O1}$ as the minimum stellar mass for which a convective core
is still present even after its central hydrogen has decreased by a significant amount
($X_{\rm c}\sim$0.2)
from the beginning of the main sequence.
$M_{\rm\,O2}$ is set equal to $M_{\rm\,O1}$+0.3$M_\odot$.
This choice is supported by
the modelling of the open cluster M\,67 (see
\citealt{Bressan_etal12}), which indicates an
overshooting efficiency $\Lambda_{\rm c}\simeq0.5$ already at
masses of M$\sim\!1.3$~\Msun\ for solar-metallicity stars;
and by the SMC cluster NGC~419 (\citealt{Girardi_etal09}, \citealt{Kamath_etal10}),
in which the turn-off probes masses between $\sim\!1.65$ and
1.9~\Msun.
\begin{figure}
  \resizebox{\hsize}{!}{
  \includegraphics{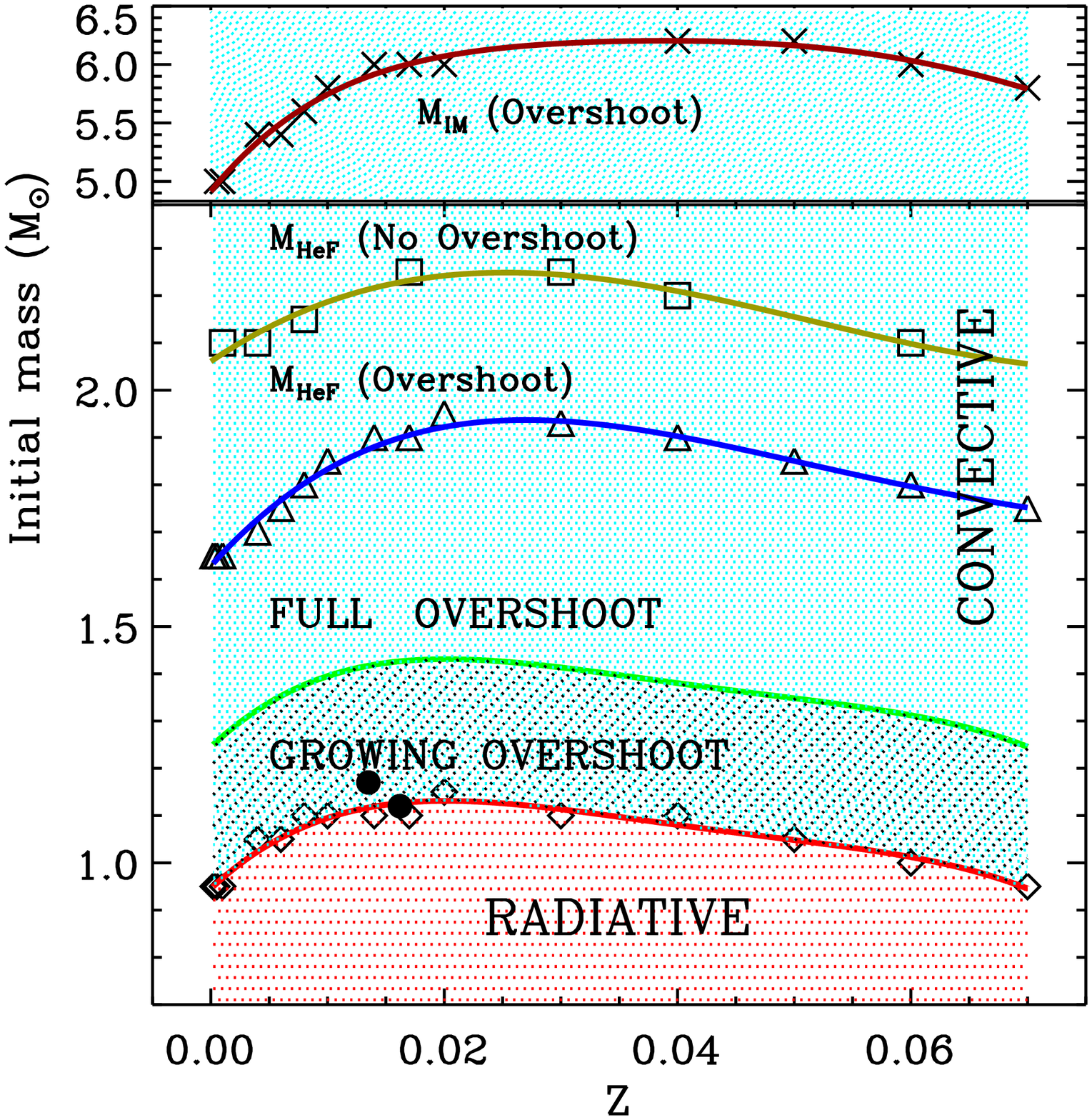}
  \includegraphics{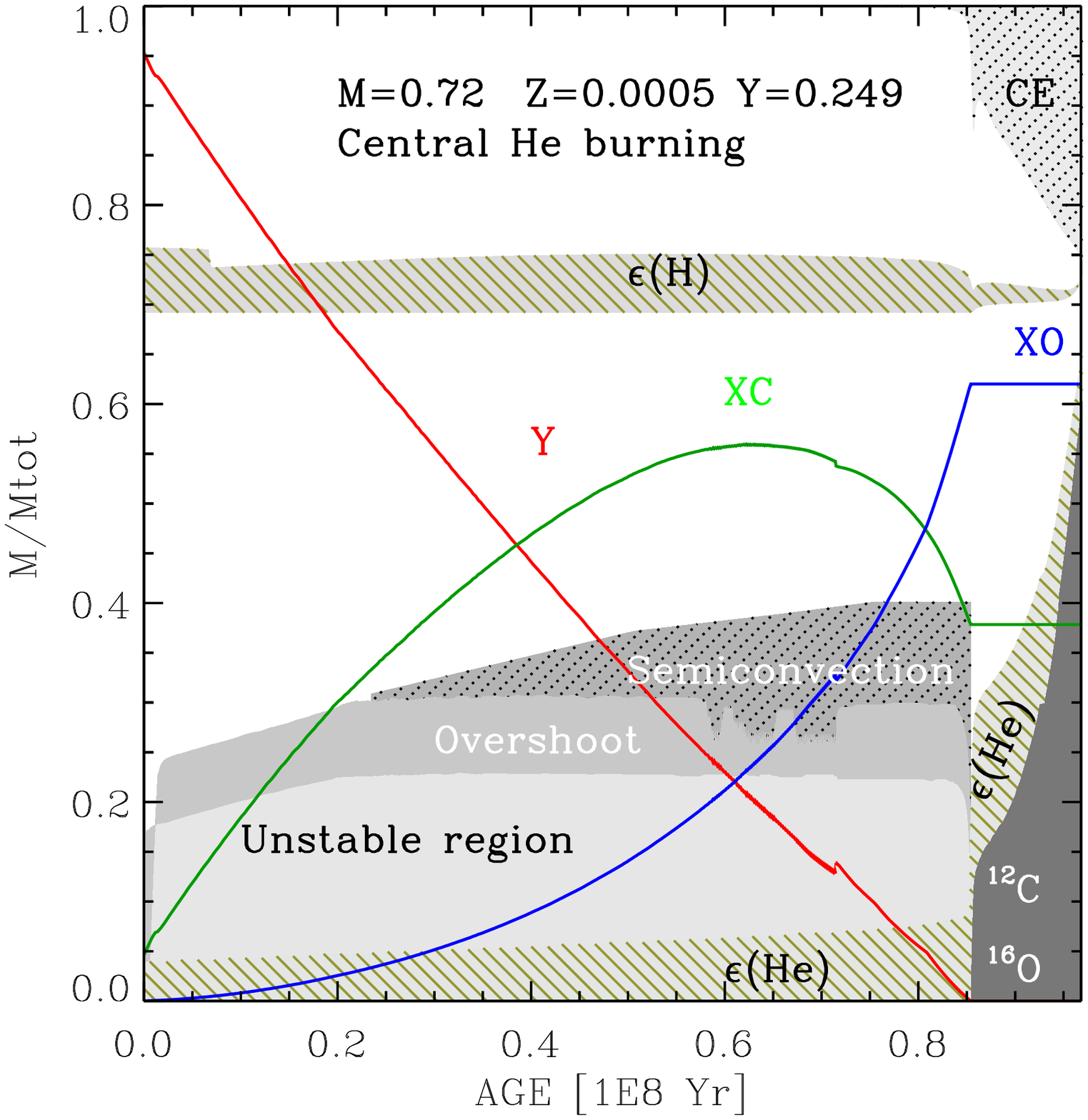}
  }
  \caption{Left: critical masses as a function of the metallicity.
  The filled dots represent the two models
that best reproduce the asteroseismic data of Dushera \citep{SilvaAguirre13}.
   Right: overshooting and semiconvective regions, during central He burning.}
    \label{fig_conv_core}
\end{figure}
The run of the critical masses $M_{\rm\,O1}$ and $M_{\rm\,O2}$
as a function of the initial metallicity are shown in the left panel of  Figure \ref{fig_conv_core}.
Constraints on these critical masses are being set by asteroseismology. Indeed
while earlier observations of $\alpha$\,Cen~A
suggest negligible overshooting in
solar-metallicity stars of mass $\sim\!1.1$~\Msun\ \citep{deMeulenaer_etal10}, recent
studies of the nearby old low-mass star HD~203608, with $[Z/X]\simeq-0.5$,  support the existence
of sizable overshoot ($\alpha_{\rm ov}=0.17$ corresponding to
$\alpha_{\rm ov}\simeq0.32$ in our formalism) at masses as low as
$0.95$~\Msun, which is probably just slightly above the $M_{\rm\,O1}$
limit at the corresponding metallicity \citep{Deheuvels_etal10}.
The full dots near $M_{\rm\,O1}$ at Z$\sim$0.015 represent the location of the two models
that best reproduce the asteroseismic data of Dushera \citep{SilvaAguirre13},
both requiring a sizeable overshoot ($\sim~0.2\,H_P$).
Effects of core overshoot during the evolution on the main sequence, already thoroughly described in literature, 
are being used to obtain indirect calibrations of this process.
Among them it is important to remind the difference between the threshold initial mass
for undergoing helium flash  between models with and without overshoot.
The critical threshold as a function of the  metallicity, M$_{HeF}$, decreases significantly
with overshoot, as indicated in Figure \ref{fig_conv_core}.
The difference in mass between the two cases is significant and is certainly larger than the
accuracy of the most recent determination of the mass with asteroseismic observations of evolved stars.
Indeed, recent detailed studies of the evolutionary properties of He burning stars
suggest to use the morphology of the red clump to trace stars with the mass around this critical transition
\citep{Girardi_MBR13}. Clearly, asteroseismology is opening a new window to constrain the mixing efficiency
in the transition region from $M_{\rm\,O1}$ to $M_{\rm\,O2}$.

\section{Overshoot \& Mixing during the He Burning Phase}
The presence of  mixing beyond the unstable region in the core of He burning stars has been suspected since
about fifty years.

{\sl Local Overshoot.} As helium is converted into carbon in the convective core of He burning stars,
the free-free opacity increases and so  the radiative temperature gradient.
This produces a discontinuity in the radiative temperature gradient just at the border of the convective core.
It has been shown that this condition is unstable because
the pollution with carbon rich material due to possible perturbations of any kind,
renders the surrounding radiative layers irreversibly unstable to convection \citep{CGR71}.
Thus, if these perturbations are allowed to occur by means of some artificial description
of this kind of  mixing, then the unstable region tend to grow during the evolution.
This effect, known as {\it local} overshoot,
increases the central He-burning lifetime  almost proportionally
to the growth of the core mass.

{\sl Semi-convection.}
In the more advanced stages, the star develops
an intermediate zone, just above the convective core, which is marginally unstable to convection.
This instability, known as semi-convection, gives rise to a smooth chemical  profile
maintaining the neutrality of the medium against convection. This mechanism is similar
to the semi-convection that appears during hydrogen burning phase of massive stars
though, in that case, it is due to the dependence of the electron scattering opacity on the
hydrogen fraction \citep{Schw58}. Notice that, in spite of the presence of a gradient in mean molecular weight,
the  Schwarzschild criterion should be preferred to the Ledoux one \citep{Kato66,Spiegel69}.
Semi-convection contributes to the mixing in the central regions and
gives rise to a further increase of the central helium exhausted core.

{\sl Breathing Pulses.}
Towards the end of central helium burning the star may undergo one or more breathing pulses of convection.
This kind of instability is due to the luminosity feedback produced by the increase of the central He content
that follows the growth of the mixed region when it enters a steep composition profile \citep{CCTP85}.
A further growth of the He exhausted core is produced by this mechanism which,
in the HR diagram of intermediate mass stars,
can be recognized by the presence of  secondary blue loops toward  central He exhaustion.

{\sl Non local overshoot.} In presence of sizable {\it non local} overshoot, the discontinuity of the
temperature gradient shifts well within the radiative stable regions
where the radiative gradient is well below the adiabatic one.
In this case the pollution of layers above the convective core does not destabilize
the surrounding stable regions and  local overshoot does not appear \citep{Bressan86}.
If one allows for large overshoot then also the semi-convective instability and even the
breathing pulse phenomenon disappear.
For the choice made in PARSEC, a sizable semi-convective region appears after the central He
mass fraction falls below about 60\% (right panel of Figure \ref{fig_conv_core}).

Overshoot (non local and/or local), semi-convection and breathing pulses may increase considerably
the amount of fuel  that the star can use during the central He-burning phase, increasing
its lifetime. But the impact they have on the following phase, the Early Asymptotic Giant branch phase,
is even larger. The star enters this phase with a larger he-exhausted core and the path toward the
double shell phase may be shortened by a significant fraction.
It has been soon recognized that the ratio between the observed number of stars in the E-AGB
and in the HB branches of globular clusters, being proportional to the lifetime ratio of the
corresponding evolutionary phases,  may constitute a strong diagnostic for the efficiency of mixing
during the helium burning phase \citep{Buonanno85}. On the other hand, this simple diagnostic may be challenged
by the presence of multiple  populations because those with high helium content populate the HB but
could escape the E-AGB phase.  Nevertheless in some metal rich clusters there
is no evidence of the latter evolutionary path and this diagnostic may still be used.
For example, in the case of 47~Tuc we obtain from the ACS HST  data \citep{Sarajedini07}
R2=$({n_{EAGB}}/{n_{HB}})$=0.14--0.15. This value is in good agreement with
the corresponding one predicted by PARSEC isochrones with  [$\alpha$/Fe]=0.4 and global metallicity
Z=0.006  and for a HB mass between 0.65~M$_\odot$ and 0.7~M$_\odot$.
\section{Overshoot at the bottom of the convective envelope}
It has been argued that a moderate overshoot region ($0.25-1.0\,H_P$) below
the base of the convective envelope
may help to better reproduce the observed location of the RGB Bump in the red giant
branch of low mass stars in globular clusters and old open clusters \citep{Alongi_etal91}.
The observed RGB bump in globular clusters is about 0.2 to 0.4~mag
fainter than that predicted by models without envelope overshooting
\citep{DiCecco_etal10} though,
at the higher metallicities, this result depends on the adopted
metallicity scale. Adopting an overshoot size of
$\Lambda_{\rm e}=0.5\,H_P$, the RGB bump becomes typically $\sim\!0.3$~mag fainter
than in models with negligible envelope overshoot \citep{Alongi_etal91},
and in good agreement with observations.
This value is in very good agreement with the overshoot size (not fully adiabatic however) at the base of the convective
envelope in the Sun, that has been estimated to be $\Lambda_{\rm e}\sim0.4-0.6\,H_P$
using solar oscillations data \citep{ChristensenDalsgaard_etal11}.
\begin{figure*}
  \resizebox{\hsize}{!}{
  \includegraphics{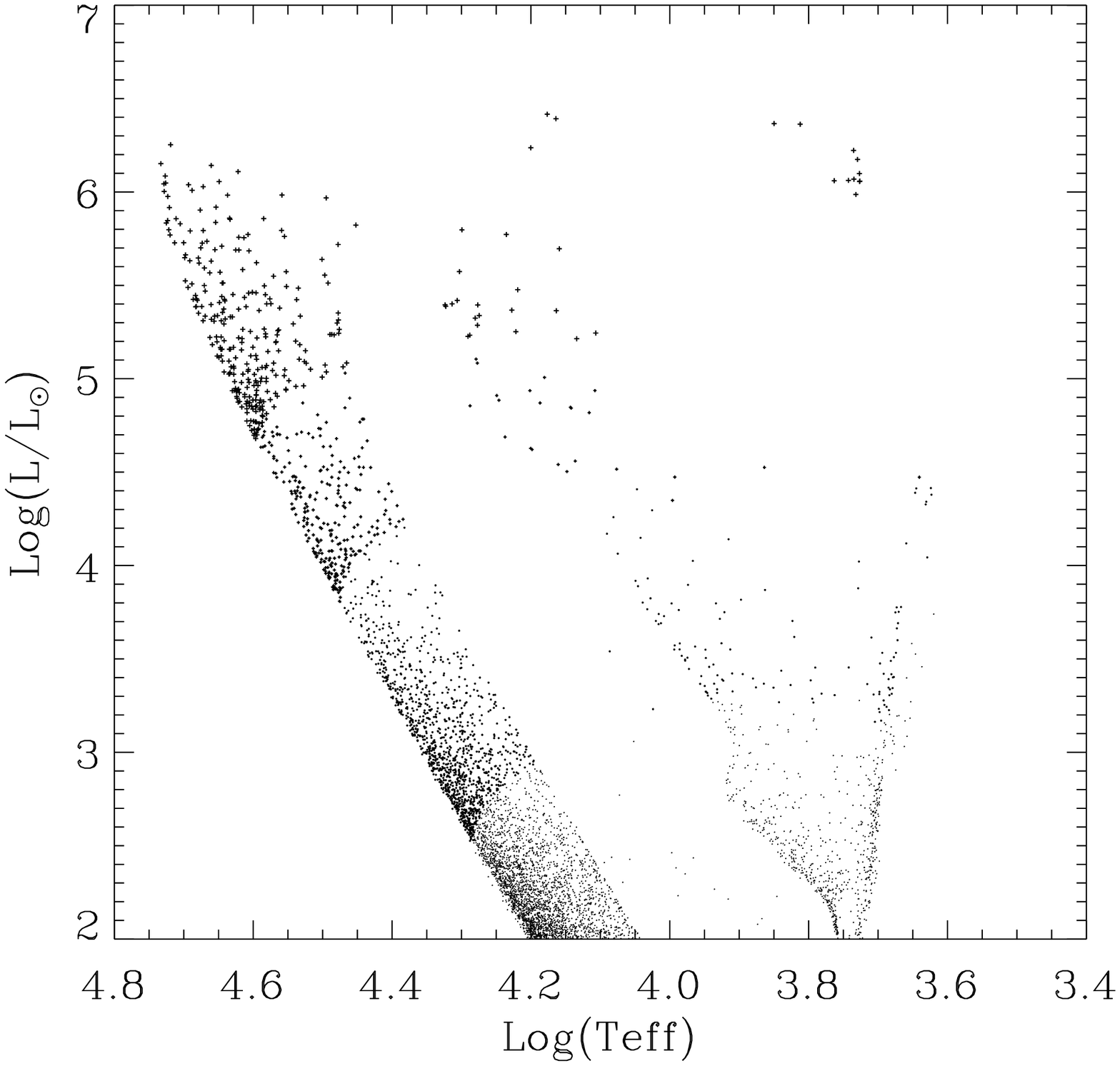}
  \includegraphics{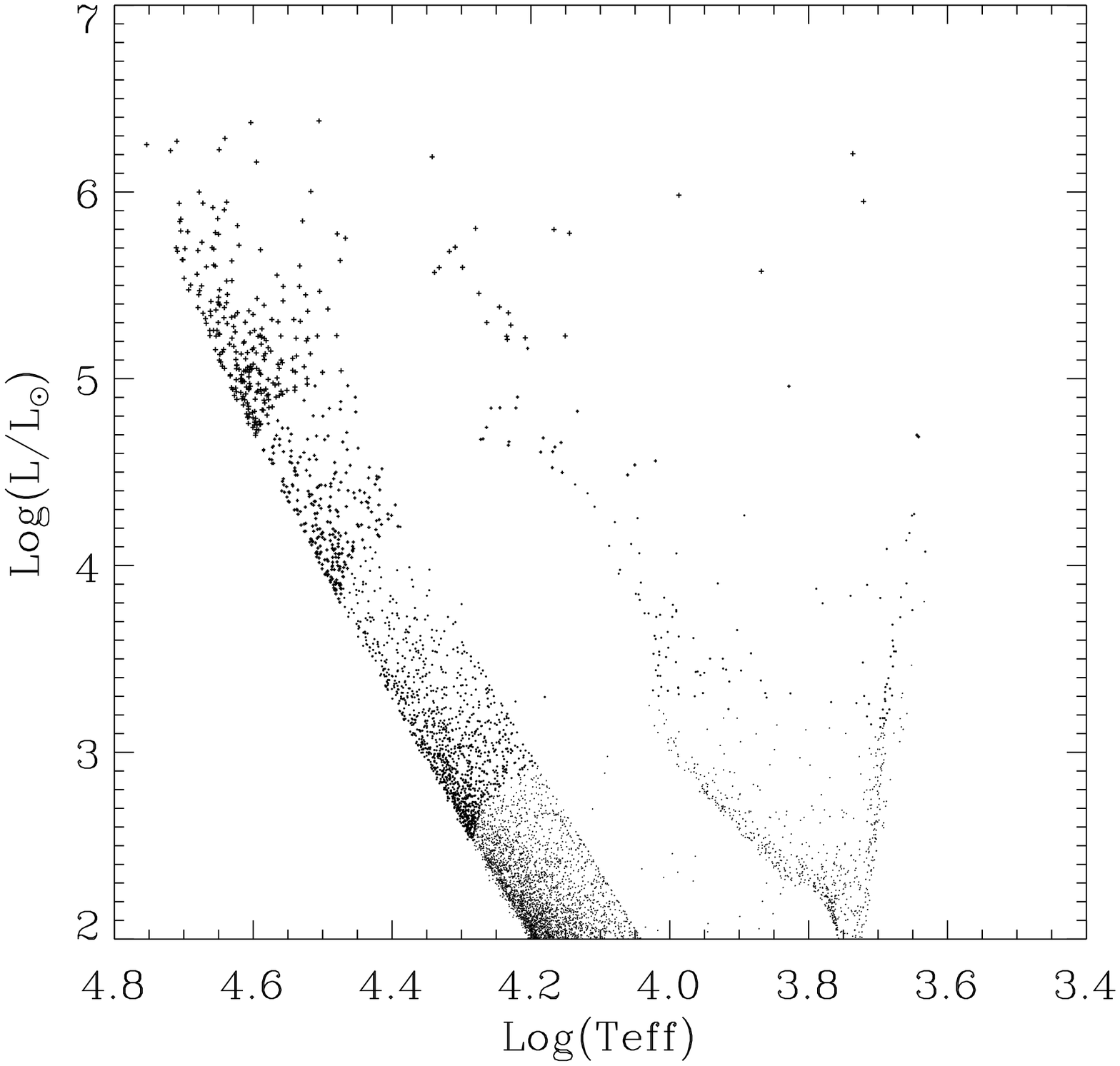}
  \includegraphics{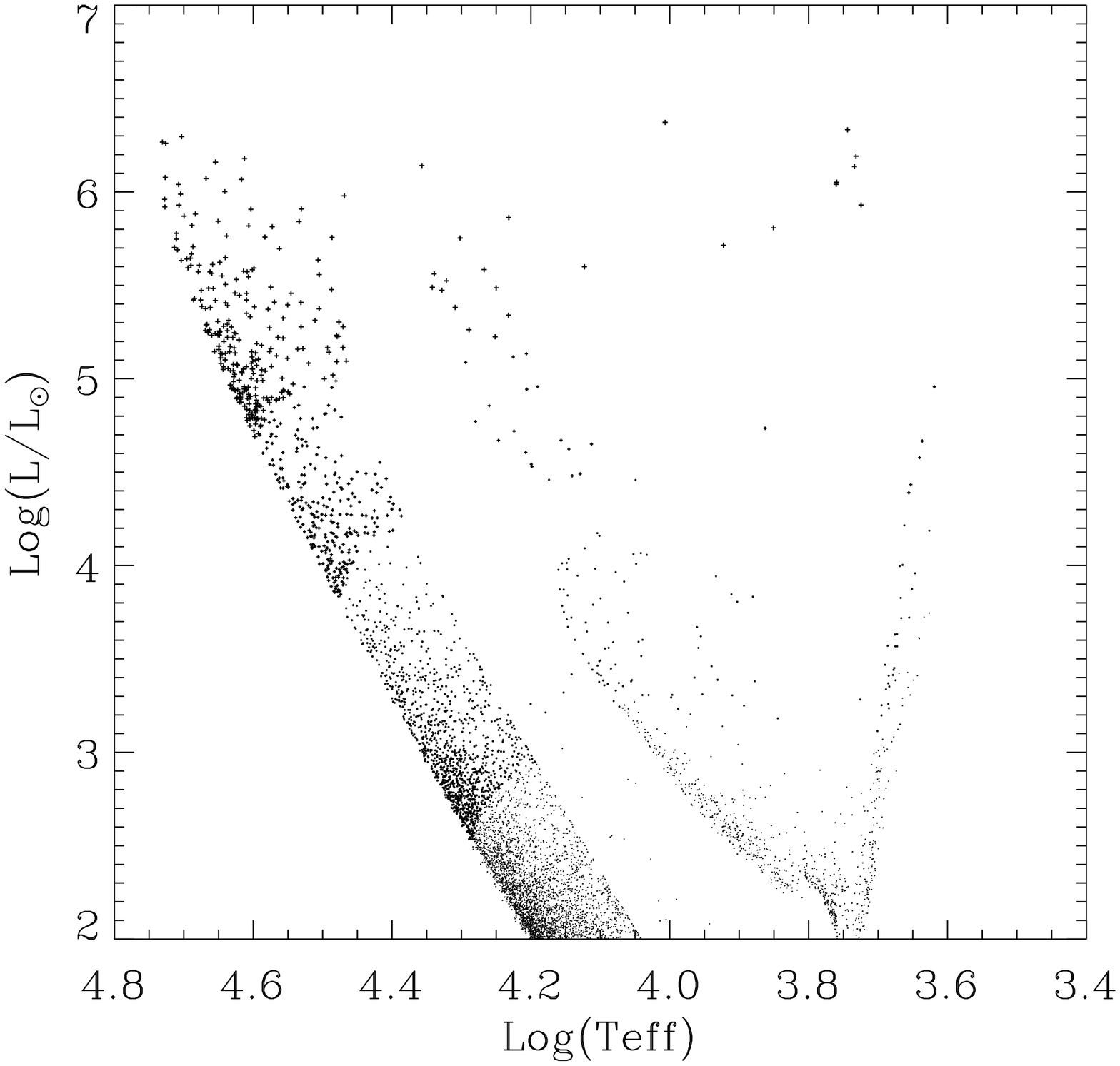}
  }
  \caption{Envelope overshoot and extension of Blue Loops in intermediate and massive stars at Z=0.001.
  From left to right the simulations are made with EO=0.7~H$_P$, 2~H$_P$ and 4~H$_P$, respectively.
  For sake of clarity, different symbol sizes and total number of stars have been considered in the mass ranges
  M$<$4M$_\odot$, 4M$_\odot$$\leq$M$<$10M$_\odot$, 10M$_\odot$$\leq$M$<$20M$_\odot$ and M$\geq$20M$_\odot$}
    \label{fig_EOHR}
\end{figure*}

Another interesting effect of the overshoot at the base of the convective envelope
concerns the extension of the blue loops of intermediate mass and massive stars.
During their central helium burning phases these
stars may perform a characteristic excursion
from the red to the blue side of the HR diagram, commonly referred to as the ``blue loop".
In nearby star-bursting dwarf  galaxies this excursion give rise to
two well resolved sequences of red helium burning (RHeB) and  blue  (BHeB) and stars \citep[e.g.,][]{McQuinn2010},
the latter being almost attached or even superimposed to the
main sequence in the more metal poor galaxies.
Indeed  the blue loops become more pronounced  at lower metallicities, but they may depend on
several other factors. In particular it has been found that their extension
decrease or they are even suppressed using either a lower
$^{12}$C($\alpha,\gamma$)$^{16}$O reaction or a larger amount of core overshoot
\citep{Alongi_etal91, Godart13}. In spite of the
strong evidence in favor of  a significant amount of core overshoot
(or an extended mixing beyond the formal convective core),
models with convective core overshoot fail to reproduce the observed
blue loops.
To cure this problem \cite{Alongi_etal91} suggested to use a certain amount of overshoot from the convective envelope.
They showed that by using an extra mixed region of at least 0.7~H$_P$ below the formal
Schwarzschild border of the envelope convection, they were able to obtain extended blue loops
even with models computed with convective core overshoot.

It is possible to check the goodness of this assumptions by comparing the models with
the colour magnitude diagrams of star-forming regions in nearby low metallicity galaxies.
These regions, that contain a large number of intermediate mass and massive stars,
are generally dominated by the latest star formation episodes for which one
may assume a narrow range in metallicity or, often, even a single metallicity.
We are analyzing this issue by comparing new evolutionary tracks with
observed C-M diagram of selected nearby dwarf galaxies from the sample of \cite{Bianchi2012} as well as \cite{Dalcanton2009}.
We show in Figure \ref{fig_EOHR} how the HR diagram of intermediate/massive stars
changes by  increasing the amount of envelope overshoot.
The simulations shown in Figure \ref{fig_EOHR} refers to a metal poor environment (Z=0.001)
and the amount of envelope overshoot is
0.7~H$_P$ (left panel) the standard choice in PARSEC, 2~H$_P$ (middle panel) and 4~H$_P$ (right panel).
A similar result is obtained for Z=0.002.
The comparisons with the observed CM diagrams of WLM, NGC~6822 and Sextans~A, three dwarf galaxies with spectroscopic metallicities
confined within the above range, indicate the need of an extra mixing of at least 2~H$_P$ below the bottom
of the convective envelope (Tang et al 2013). With the standard PARSEC choice (0.7~H$_P$), the C-M diagram of
these galaxies could be reproduced only adopting a significantly lower metallicity, Z=0.0005.
Though large, the above value is not uncommon and similar values are used to
enhance the efficiency of the carbon  dredge-up during the thermally pulsing Asymptotic Giant Branch phase \citep{Kamath12}.

\section{Conclusions}
Convective mixing plays a major role in stellar evolution, since it can modify the structure of the star  in an irreversible way.
Whether convection is accompanied by significant overshoot  remains an unsolved problem.
The existence of extra mixing is one of the most uncertain factors in stellar astrophysics affecting
H-burning lifetimes and the lifetimes of advanced evolutionary phases (HB, E-AGB),
luminosities and effective temperatures (Main Sequence termination, Clump stars, Blue He burning stars)
and in particular cases also the surface chemistry.
Classical tests with colour magnitude diagrams indicate the presence of this extra mixing outside the central convective region.
Recent asteroseismic observations put even more robust constraints on this phenomenon, even if
the nature of this mixing remains still unclear since it could also be due to the effects of rotational mixing.
In some cases observations are even more challenging.
In fact if one allows for a larger mixing above the convective core during the H-burning phase,
the theoretical blue loops of intermediate and massive stars become less extended, at variance with
observations, unless an even higher extra mixing is applied to the bottom of the convective envelopes.
The size  required to reproduce
the observed blue loops in the CM diagrams of well studied star forming regions in nearby dwarf galaxies ($\geq~2\,H_P$),
is significantly larger than that required to reproduce the RGB bump in globular clusters ($\sim~0.5\,H_P$).

%
%

%


\end{document}